\documentclass{article}
\usepackage{spconf,amsmath,graphicx}
\usepackage{xcolor}
\usepackage{multirow}
\usepackage{amsmath}
\usepackage[linesnumbered,ruled]{algorithm2e}
\usepackage{float}
\usepackage[numbers,sort&compress,square]{natbib}
\usepackage{amssymb}
\usepackage{caption}
\usepackage{subcaption}
\usepackage{hyperref}

\def\rset{\mathbb{R}}
\def\cset{\mathbb{C}}

\title{Diagonal State Space Augmented Transformers for Speech Recognition}
\name{George Saon, Ankit Gupta and Xiaodong Cui}
\address{IBM Research AI, Yorktown Heights, USA}
\begin{document}
\ninept
\maketitle
\begin{abstract}
  We improve on the popular conformer architecture by replacing the
  depthwise temporal convolutions with diagonal state space (DSS)
  models. DSS is a recently introduced variant of linear RNNs obtained
  by discretizing a linear dynamical system with a diagonal state
  transition matrix. DSS layers project
  the input sequence onto a space of orthogonal polynomials where the choice of
  basis functions, metric and support is controlled by the eigenvalues
  of the transition matrix. We compare neural transducers with either
  conformer or our proposed DSS-augmented transformer (DSSformer) encoders on three
  public corpora: Switchboard English conversational telephone
  speech 300 hours, Switchboard+Fisher 2000 hours, and a
  spoken archive of holocaust survivor testimonials called MALACH 176 hours. On
  Switchboard 300/2000 hours, we reach a single model performance of
  8.9\%/6.7\% WER on the combined
  test set of the Hub5 2000 evaluation, respectively, and on MALACH we improve the WER by 7\% relative over the
  previous best published result. In addition, we present empirical
  evidence suggesting that DSS layers learn damped Fourier basis functions
  where the attenuation coefficients are layer specific whereas the
  frequency coefficients converge to almost identical linearly-spaced values across all layers.
\end{abstract}

\section{Introduction and related work}

An interesting alternative to the ubiquitous transformer
architecture is the recently introduced structured
state space sequence model (S4) which showed promising
results for modeling long range dependencies on the LRA (Long Range
Arena) benchmark for sequence-level classification of different modalities such as text, images and mathematical expressions~\cite{gu2021efficiently}. The main
idea behind S4 is that the input sequence can be modeled as a linear
RNN obtained by discretizing a continuous state space model. The
physical meaning of a state in S4 is a time-varying vector of linear
expansion coefficients used to approximate the input sequence with
orthogonal polynomials under a given measure and support (weighting function and input window)~\cite{gu2020hippo}. The appeal of these
models is that they can be efficiently implemented as full sequence
convolutions running in ${\cal O}(T\log T)$ instead of the ${\cal O}(T^2)$ complexity for self-attention with $T$ being the input sequence length. Moreover, these models are solidly grounded in function approximation theory and have interpretable parameters in terms of basis functions, measures and time sampling intervals. 

In~\cite{gu2021efficiently} the authors
consider a diagonal plus low-rank approximation of the state
transition matrix which simplifies the convolutional kernel
estimation. In~\cite{gupta2022diagonal}, the authors observed that there is no
loss in performance when assuming that the transition matrix is
diagonal with complex eigenvalues which is conceptually simpler and
straightforward to implement compared to~\cite{gu2021efficiently}. Because of this, diagonal state space (DSS) models will be adopted in this paper. In both
works, the authors initialize the diagonal entries of the state
transition matrix with the eigenvalues of a higher-order polynomial
projection operator (HiPPO) matrix such that the input function is
uniformly approximated with Legendre polynomials over a sliding window
of fixed length.  In~\cite{mehta2022long} the authors argue that parameterizing the eigenvalues in log-space and initializing them with -exp for the real parts and +exp for the imaginary parts is just as effective and improve the
DSS model further by augmenting it with self-attention to better
capture local dependencies. In~\cite{gu2022parameterization}, the authors revisit the parameterization and initialization of DSS and propose eigenvalue initialization schemes with constant negative real parts with respect to the eigenvalue index and imaginary parts which scale either inversely or linearly with the eigenvalue index. The former results in projecting the input onto the space of Legendre polynomials with uniform weighting from the beginning of the sequence up to the current time whereas the latter amounts to using damped Fourier basis functions as approximators with an exponentially decaying weighted history.

While DSS has been primarily developed as an alternative to
self-attention, the dual RNN/convolutional representation suggests
that it has potential to outperform the depthwise temporal
convolutions in the conformer architecture~\cite{gulati2020conformer}. We echo the
findings of~\cite{mehta2022long} which indicate that self-attention and
DSS exhibit complementary behaviour and do not necessarily subsume each other. Given the popularity and effectiveness of conformers for both hybrid~\cite{zeineldeen2022improving} and end-to-end ASR~\cite{tuske2021limit,sainath2022improving,li2022recent,shi2022streaming,zhang2022bigssl}, several other avenues
have been explored in the literature to either improve the conformer
architecture or the training recipe. In~\cite{burchi2021efficient}, the authors
use grouped self-attention and progressive down-sampling to reduce the
complexity of the self-attention layer. In~\cite{guo2021recent}, the
authors provide training recipes and extensive comparisons between
conformers and transformers on several corpora. In~\cite{wang2020efficient} the
authors replace the transformer layer with performer. In~\cite{li2021efficient},
the authors use linear self-attention layers. In~\cite{zeineldeen2022improving} the
authors use two convolutional layers for each conformer block and layer normalization instead of batch norm. Similar to our work, in~\cite{jiang2022nextformer},
the authors replace the convolutional layers with a more powerful representation called ConvNeXt.

The main contributions of this work are summarized below:
\begin{itemize}
\setlength\itemsep{-1pt} % default value: 6pt
\item We apply diagonal state space models to speech recognition and report experimental results on three public corpora.
\item We show that DSSformers outperform conformers when used as encoders for neural transducers and achieve state-of-the-art results for single non-AED models on Switchboard telephony speech and MALACH.
\item We study the effect of DSS initialization and provide some insights into what the DSS layers actually learn.
\end{itemize}
The rest of the paper is organized as follows: in section~\ref{dss-sec} we review the DSS formalism; in section~\ref{exp-sec} we present experimental evidence of its utility and in section~\ref{con-sec} we summarize our findings.

\section{DSS formulation}
\label{dss-sec}
We briefly review the main concepts behind the diagonal state spaces framework for readers from the ASR community who may not be familiar with this new sequence-to-sequence modeling approach.\\ 
\subsection{State space model} 
Borrowing some definitions and notations  from~\cite{gu2021efficiently,gupta2022diagonal}, a continuous state space model (SSM), sometimes referred to in the literature as a linear time-invariant or a linear dynamical system, is defined by the linear ODE: 
\begin{align}
\nonumber x^\prime(t)&={\bf A}x(t)+{\bf B}u(t), &{\bf A}\in\rset^{N\times N},~{\bf B}\in\rset^{N\times 1} \\
y(t)&={\bf C}x(t),& {\bf C}\in\rset^{1\times N}\label{ssm}
\end{align}
~~\\
that maps the continuous 1-dimensional input $u(t)\in\rset$ to an $N$-dimensional latent state $x(t)\in\rset^{N}$ before projecting it to a 1-dimensional output $y(t)\in\rset$. The state space is parameterized by the state transition matrix ${\bf A}$ as well as trainable parameters ${\bf B},{\bf C}$.

\subsection{Discretization and link to linear RNNs} Consider a sampling interval $\Delta>0$ and define  $u_k:=u(k\Delta), k=0\ldots L-1$ the sampled input signal. Correspondingly, we have $x_k=x(k\Delta)$ and $y_k=y(k\Delta)$. Equation~(\ref{ssm}) can be turned into a discrete recurrence that maps $(u_0,\ldots,u_{L-1})\mapsto(y_0,\ldots,y_{L-1})$ by integrating over $[(k-1)\Delta,k\Delta]$ under the zero-order hold (ZOH) assumption $u(t)=u_k,~(k-1)\Delta\le t< k\Delta$:
\begin{align}
\nonumber x_k&=\overline{\bf A}x_{k-1}+\overline{\bf B}u_k, &\overline{\bf A}=e^{{\bf A}\Delta},~\overline{\bf B}=(e^{{\bf A}\Delta}-{\bf I}){\bf A}^{-1}{\bf B}\\
y_k&=\overline{\bf C}x_k, &\overline{\bf C}={\bf C}\label{rnn}
\end{align}

\subsection{Convolutional representation} With the convention $x_{-1}=0$, the recurrence can be unrolled and rewritten by eliminating the state variables $x_k$:
\begin{equation}
   y_k = \sum_{j=0}^k\overline{\bf C}\overline{\bf A}^j\overline{\bf B}u_{k-j},\qquad k=0,\ldots,L-1
   \label{unroll}
\end{equation}\\
By grouping the scalar coefficients $\overline{\bf C}\overline{\bf A}^k\overline{\bf B}$ into the SSM kernel $\overline{\bf K}\in\rset^L,~
\overline{\bf K}=(\overline{\bf C}\overline{\bf B},\overline{\bf C}\overline{\bf A}\overline{\bf B},\ldots,\overline{\bf C}\overline{\bf A}^{L-1}\overline{\bf B})$, (\ref{unroll}) can be elegantly reformulated as a convolution
\begin{equation}
y=\overline{\bf K}*u
\label{conv}
\end{equation}\\
Computing~(\ref{conv}) naively would require ${\cal O}(L^2)$ operations. Instead, we observe that $y_k$ is the coefficient of $z^k$ of the product $u(z)\cdot\overline{K}(z)$ of two $(L-1)$-degree univariate polynomials $u(z)=\sum_{i=0}^{L-1}u_i z^i$ and $\overline{K}(z)=\sum_{i=0}^{L-1}\overline{K}_i z^i$. By the circular convolution theorem, this product can be computed efficiently in ${\cal O}(L\log L)$ using FFT and its inverse.

\subsection{Diagonal state spaces} 
Based on the above, computing $y$ from $\overline{\bf K}$ and $u$ is easy; the hard part is how to compute $\overline{\bf K}$ efficiently. The main result in~\cite{gupta2022diagonal} states that if ${\bf A}\in\cset^{N\times N}$ is diagonalizable over $\cset$ with eigenvalues $\lambda_1,\ldots,\lambda_N$ such that, $\forall i$, $\lambda_i\ne 0$ and $e^{L\lambda_i\Delta} \ne 1$, there $\exists w \in \cset^{1\times N}$ such that
\begin{equation}
    \overline{\bf K} = w\cdot\mathbf{\Lambda}^{-1}\cdot\mbox{row-softmax}({\bf P})
\label{dss}
\end{equation}
~~\\
where ${\bf P}\in\cset^{N\times L},~p_{ik}=\lambda_i k\Delta$ and $\mathbf{\Lambda}=\mbox{diag}(\lambda_1,\ldots,\lambda_N)$. The proof uses the diagonalization of ${\bf A}={\bf V}\mathbf{\Lambda}{\bf V}^{-1}$ which, from the expression of $\overline{\bf A}$ from~(\ref{rnn}), implies $\overline{\bf A}^k = e^{k{\bf A}\Delta}={\bf V}e^{k\mathbf{\Lambda}\Delta}{\bf V}^{-1}$, and the geometric series identity $\sum_{k=0}^{L-1}z^k=\frac{z^L-1}{z-1}$. We refer the reader to~\cite{gupta2022diagonal} for the complete proof.

\subsection{DSS layer} 
A DSS layer operates as follows. It receives an $H\times L$ input sequence and produces an $H\times L$ output sequence where $H$ is the number of channels and $L$ is the sequence length. It does this by applying $H$ DSS kernels to the input (with a shortcut connection) according to (\ref{conv}), one for each coordinate. We apply a Gaussian Error Linear Unit (GELU) nonlinearity to the result followed by an $H\times H$ pointwise linear layer needed to exchange information between the dimensions. After mixing, we apply a Gated Linear Unit (GLU) activation to the output. The implementation of a DSS layer as described so far is publicly available at \url{https://github.com/ag1988/dss}.

For a state space dimension $N$, the trainable parameters of the DSS layer are: $\mathbf{\Lambda}_{re},\mathbf{\Lambda}_{im}\in\rset^N$ the diagonal entries of the transition matrix (tied across all channels), $W\in\cset^{H\times N}$ from~(\ref{dss}), $\Delta\in\rset^H$ the time sampling intervals, and $W_{out}\in\rset^{H\times H}$ the output mixing matrix.

\begin{figure}[H]
    \centering
    \includegraphics[width=0.5\textwidth]{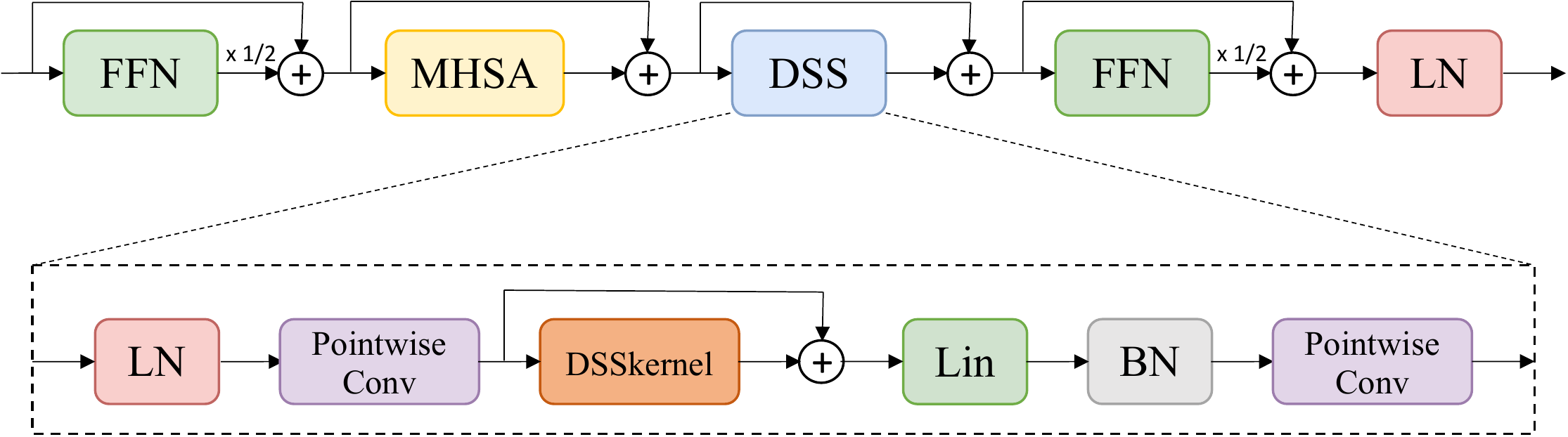}
    \caption{Proposed architecture: top DSSformer block, bottom DSS module (non-linearities are omitted for clarity).}
        \label{dss-arch}
\end{figure}

Just like the depthwise separable convolution module in the conformer architecture, the DSS layer is sandwiched between two pointwise convolutions which serve to increase the inner dimension (typically by a factor of 2) on which the layer operates as shown in Figure~\ref{dss-arch}.

\section{Experiments and results}
\label{exp-sec}
We investigate the effectiveness of the proposed model on three public corpora: Switchboard English conversational telephone speech 300 hours, Switchboard+Fisher 2000 hours, and MALACH 176 hours.

\subsection{Experiments on Switchboard 300 hours}
\label{swb300}
The acoustic training data comprises 300 hours of English telephone conversations
between two strangers on a preassigned topic. We follow the Kaldi {\tt s5c} recipe~\cite{povey11} for data preparation and segmentation and report results on the
Hub5 2000 (Switchboard and CallHome), Hub5 2001 and RT'03 test sets which are processed according to the LDC segmentation and scored using Kaldi WER measurement.

\subsubsection{Feature processing}
Our feature extraction and training recipe largely mirrors~\cite{saon2021advancing} with some notable differences. We extract 40-dimensional speaker independent log-Mel features every 10ms with speaker-based mean and variance normalization augmented with $\Delta$ and $\Delta\Delta$ coefficients. We perform temporal subsampling by a factor of 2 by stacking every two consecutive frames and skipping every second stacked frame which results in 50 240-dimensional feature vectors per second. Unlike~\cite{saon2021advancing,cui2022improving,zeineldeen2022improving}, we do not use appended i-vectors as we found them to be less effective with conformer transducers.
We create 4 additional replicas of the training data using speed and tempo perturbation~\cite{ko15} both with values in $\{0.9,1.1\}$ which, together with the original data, amounts to 1500 hours of training data every epoch. We perturb the data in three different ways: (i) sequence noise injection adds, with probability 0.8, a down-scaled spectrum of a random utterance to the current utterance~\cite{saon19a}; (ii) SpecAugment randomly masks blocks in both time and frequency with the settings from~\cite{park19}; (iii) Length perturbation randomly deletes and inserts contiguous frames with probability 0.7~\cite{cui2022improving}.

\subsubsection{Transducer architecture}
We trained neural transducers (or RNN-Ts\footnote{Both terms are used interchangeably in the literature even for models where the encoder is not an RNN.})~\cite{graves12} with either conformer or DSSformer encoders with 12 layers, feed-forward dimension of 384 and 6$\times$96-dimensional attention heads for an inner dimension of 512. All DSS layers use bidirectional kernels with a state space dimension $N$=96. The joint network projects the 384-dim vectors from the last encoder layer to 256 and multiplies the result elementwise~\cite{saon2021advancing,zhang2022improving} with a 256-dim projection of a label embedding computed by a unidirectional 1024-cell LSTM prediction network. After the application of hyperbolic tangent, the output is projected to 46 logits followed by a softmax
layer corresponding to 45 characters plus BLANK. The baseline conformer RNN-T has an optimal size of 63M parameters and the DSSformer RNN-T has 73M parameters. 

\subsubsection{Training and decoding}
The models were trained in Pytorch to minimize the RNN-T loss with CTC loss smoothing from the encoder with a weight of 0.1. Training was carried out on single A100 GPUs for 24 epochs with AdamW SGD and a one cycle learning rate policy which ramps up the step size linearly from 5e-5 to 5e-4 for the first 8 epochs followed by a linear annealing phase to 0 for the remaining 16 epochs. All experiments use a batch size of 64 utterances. Decoding was done using alignment-length synchronous beam search~\cite{saon2020alignment}. We also report results with density ratio shallow language model fusion~\cite{mcdermott19} where the target LM is a 12-layer, 512-dimensional transformerXL character LM~\cite{dai2019transformer} trained on the Switchboard+Fisher acoustic transcripts (126M characters) and the source LM has the same configuration as the prediction network and was trained on the 300 hours transcripts only (15M characters). 

\subsubsection{DSS layer initialization and recognition results}
In Table~\ref{swb300-tab}, we compare the performance of baseline conformer and DSSformer transducers with different initializations of the $\mathbf{\Lambda}$ matrix. Concretely, $\mathbf{\Lambda}$ HiPPO uses the top $N$ eigenvalues with positive imaginary part from the skew-symmetric $2N\times 2N$ matrix $a_{ij}=\left\{\begin{array}{ll}2(i+1)^{1/2}(2j+1)^{1/2},&i<j\\-1/2,&i=j\\-2(i+1)^{1/2}(2j+1)^{1/2},&i>j\\ \end{array}\right.$~\cite{gupta2022diagonal}. For $\mathbf{\Lambda}$ exp random, $\lambda_n=-e^{a_n}+i\cdot e^{b_n}$ where $a_n,b_n\sim{\cal U}[-1,1]$~\cite{mehta2022long}. For $\mathbf{\Lambda}$ S4D-Inv, $\lambda_n=-\frac{1}{2}+i \frac{N}{\pi}\left(\frac{N}{2n+1}-1\right)$, whereas for $\mathbf{\Lambda}$ S4D-Lin, $\lambda_n=-\frac{1}{2}+i\pi n$~\cite{gu2022parameterization}. For all experiments, $\Delta$ is parameterized in log-space with values drawn from ${\cal U}[\log(0.001),\log(0.1)]$ and the real and imaginary parts for $w$ in~(\ref{dss}) are initialized from ${\cal N}(0,1)$. 

\begin{table}[H]
\begin{center}
\setlength\tabcolsep{6pt} % default value: 6pt
\begin{tabular}{|l|c|c|c|c|c|} \hline
\multirow{2}{*}{Encoder} & \multicolumn{3}{|c|}{Hub5'00} & \multirow{2}{*}{Hub5'01} & \multirow{2}{*}{RT'03}\\ \cline{2-4}
         & swb  & ch    & avg  &      &    \\ \hline
Conformer & 7.5  & 15.0  & 11.2 & 11.2 & 14.4\\ 
~~~~~~~-MHSA+DSS      & 8.0  & 15.9  & 12.0 & 12.2   & 15.7\\ \hline
$\mathbf{\Lambda}$ HiPPO~\cite{gupta2022diagonal} & 7.2 & 14.2 & 10.7 & 10.7 & 13.5 \\ \hline
$\mathbf{\Lambda}$ exp random~\cite{mehta2022long} & 7.3 & 14.5 & 10.9 & 10.8 & 13.4\\ \hline  
$\mathbf{\Lambda}$ S4D-Inv~\cite{gu2022parameterization} & 7.5 & 14.7 & 11.1 & 10.9 & 13.8\\ \hline
$\mathbf{\Lambda}$ S4D-Lin~\cite{gu2022parameterization} & 7.2 & 14.3 & 10.8 & 10.9 & 13.3\\ \hline
$\lambda_n=-1+i\cdot n$ & 7.1 & 13.9 & 10.5 & 10.5 & 13.3\\ \hline
\end{tabular}
\end{center}
\caption{\label{swb300-tab} Recognition results for conformer, conformer with MHSA replaced by DSS, and DSSformer transducers with different $\mathbf{\Lambda}$ initializations on Switchboard 300 hours (Hub5'00, Hub5'01, RT'03 test sets). All models are trained for 24 epochs without length perturbation and decodings are done without external LM.}
\end{table}

\begin{figure}[H]
     \hspace*{-3mm}
     \begin{subfigure}[b]{0.23\textwidth}
         \centering
         \includegraphics[width=1.1\textwidth]{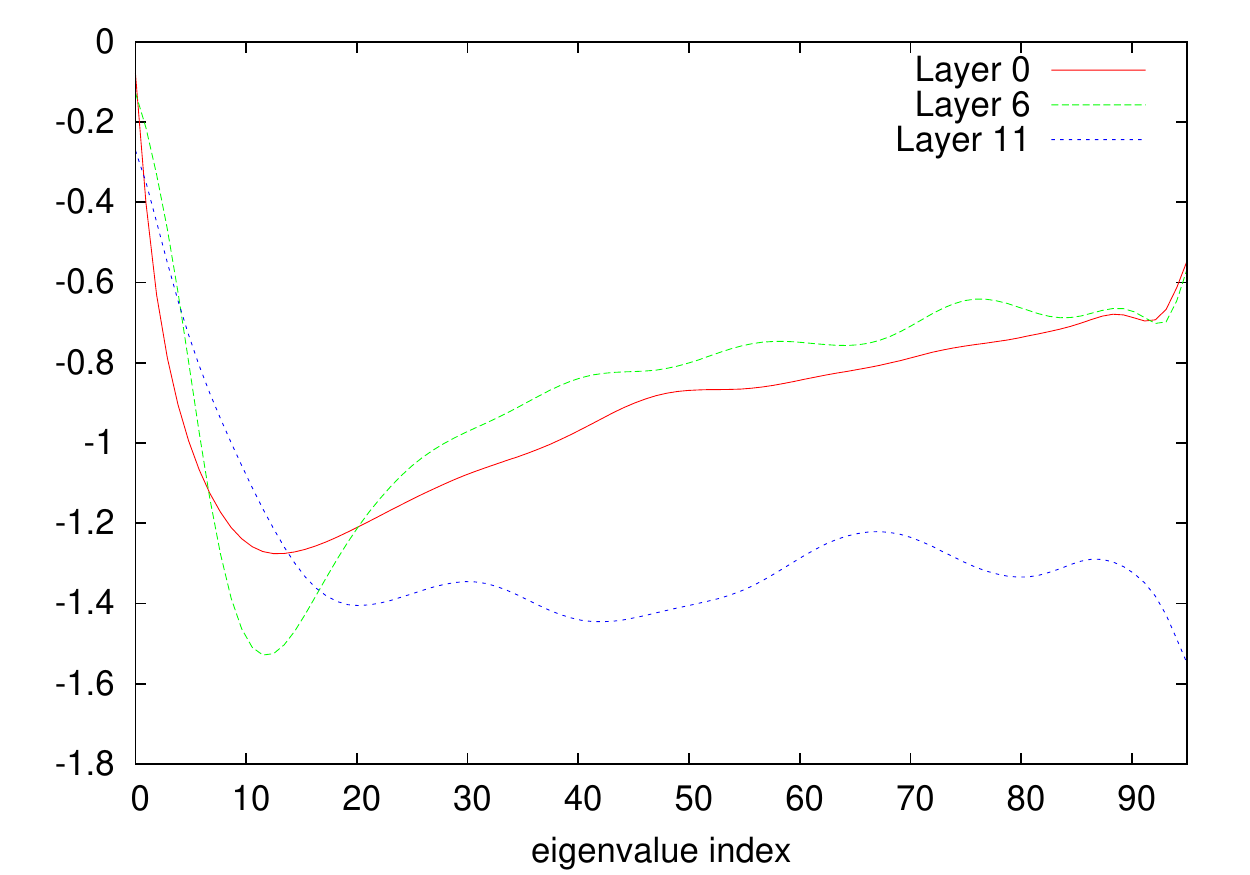}
         \caption{Real parts of trained $\lambda_n$}
         \label{re-lambda}
     \end{subfigure}
     \hspace*{2mm}
     \begin{subfigure}[b]{0.23\textwidth}
         \centering
         \includegraphics[width=1.1\textwidth]{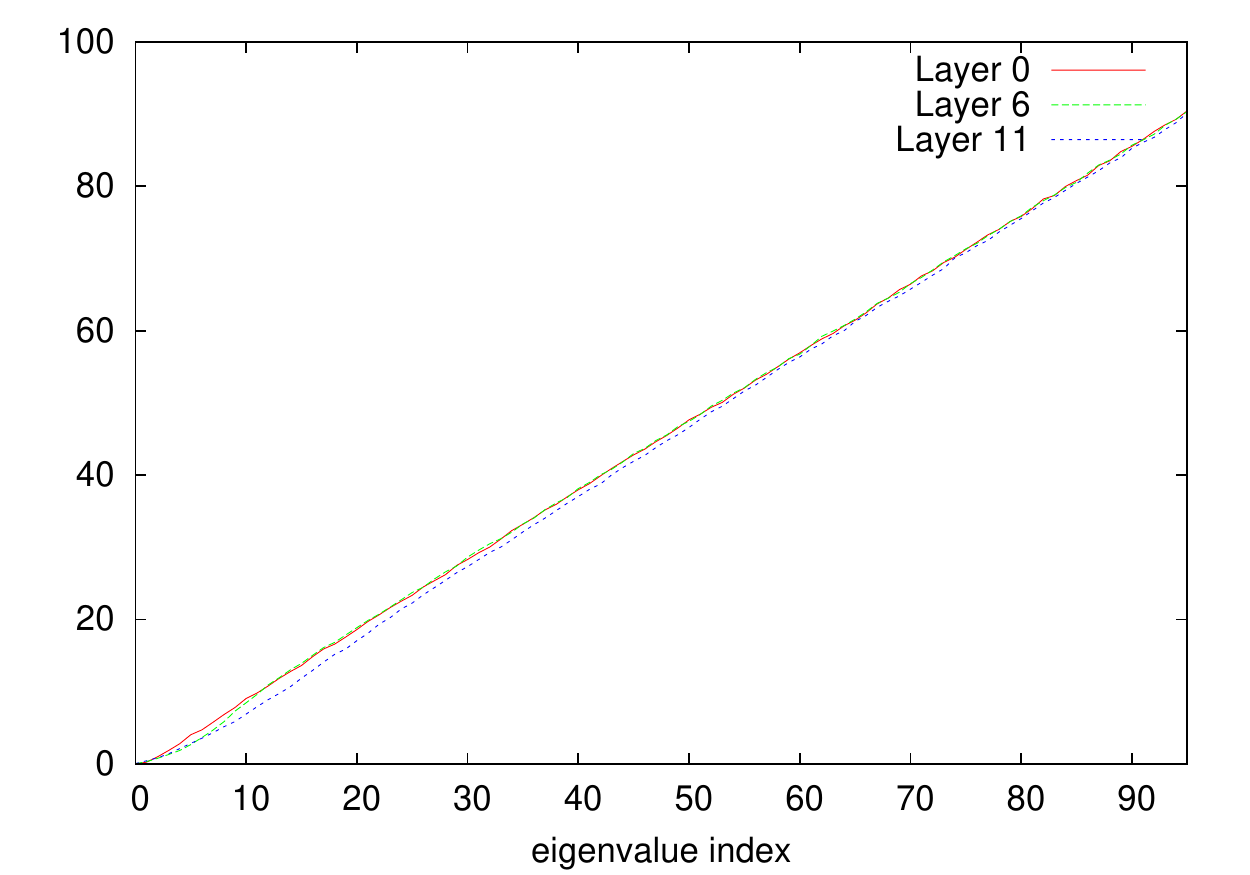}
         \caption{Imaginary parts of trained $\lambda_n$}
         \label{im-lambda}
     \end{subfigure}
    \caption{Converged eigenvalues for S4D-Lin initialization for first, middle and last layers on Switchboard 300 hours.}
        \label{lambdas}
\end{figure}

The $\mathbf{\Lambda}$ initialization from the last row in Table~\ref{swb300-tab} was motivated by inspecting the converged values of $\lambda_n$ when the DSS layers were initialized with S4D-Lin. Interestingly, the imaginary parts of $\lambda_n$ converge from $\pi n$ to approximately $0.95n$ across all layers as shown in Figure~\ref{im-lambda}. In contrast, in Figure~\ref{re-lambda} the real parts converge to values that are layer-dependent\footnote{The curves have been smoothed with Bezier interpolation for ease of visualization.}. This suggests that the DSS layers learn damped Fourier basis functions $F_n(t)=e^{-\lambda_n t}$ where the attenuation coefficients are layer specific and the frequency coefficients are linearly spaced and common across layers. The benefit of using FFT layers for mixing input sequences has also been shown in the FNet architecture~\cite{lee2021fnet}.

In Table~\ref{swb300-best} we compare the performance of our best single DSSformer model with existing approaches from the literature. Here, the model was trained for 30 epochs with length perturbation with the following settings from~\cite{cui2022improving}: insertion and deletion probabilities of 0.7, 10\% of frames selected as starting points for both, maximum deletion length of 7 frames and maximum insertion length of 3 frames. Length perturbation is lifted after 25 epochs.

\begin{table}[H]
\setlength\tabcolsep{3pt} % default value: 6pt
\begin{center}
\begin{tabular}{|l|l|l|c|c|c|c|c|} \hline
\multirow{2}{*}{Work} &\multirow{2}{*}{Model} & \multirow{2}{*}{Encoder} & \multirow{2}{*}{LM} & \multicolumn{3}{|c|}{Hub5'00} & \multirow{2}{*}{Hub5'01}\\ \cline{5-7}
                                         &                        &                          &   & swb  & ch    & avg  & \\ \hline
\cite{guo2021recent} & AED & Conformer & --   & 7.1 & 15.0 & 11.1 & -- \\ \hline
\multirow{3}{*}{\cite{tuske2021limit}} & \multirow{3}{*}{AED} & \multirow{3}{*}{Conformer} & --   & 6.7 & 13.0 & 9.9 & 10.0\\
                                         &                        &                   & LSTM$^*$ & 5.7 & 11.4 & 8.6 & 8.5\\ 
                                         &                        &                   &+Trafo$^*$ & 5.5 & 11.2 & 8.4 & 8.5\\ \hline
\multirow{2}{*}{\cite{zeineldeen2022improving}} & \multirow{2}{*}{HMM} & \multirow{2}{*}{Conformer} & n-gram & 7.1 & 13.5 & 10.3 & 10.4\\
                                         &                        &                          & Trafo & 6.3 & 12.1 & 9.2 & 9.3\\ \hline
\multirow{2}{*}{\cite{cui2022improving}} & \multirow{2}{*}{RNN-T} & \multirow{2}{*}{LSTM}    & -- & 6.9 & 14.5 & 10.7 & 11.2\\
                                         &                        &                          & LSTM & 5.9 & 12.5 & 9.2  & 9.4\\ \hline
\multirow{2}{*}{\cite{zhou2022efficient}} & \multirow{2}{*}{RNN-T} & \multirow{2}{*}{Conformer} & n-gram & -- & -- & 10.3 & 10.6\\
                                         &                       &                        & Trafo & -- & -- & 9.3 & 9.4\\ \hline
\multirow{2}{*}{Ours}                    & \multirow{2}{*}{RNN-T} & \multirow{2}{*}{DSSformer} & -- & 6.7 & 13.4 & 10.0 & 10.3\\ 
                                         &                        &                          & Trafo & 5.6 & 12.2 & 8.9  & 9.0 \\ \hline
\end{tabular}
\end{center}
\caption{\label{swb300-best} Performance comparison of DSSformer transducer with other single-model approaches from the literature on Switchboard 300 hours ($^*$ are cross-utterance LMs).}
\end{table}

\subsection{Experiments on Switchboard+Fisher 2000 hours}
\label{swb2000}
The second set of experiments was carried out on 1975 hours (9875 hours after augmentation) comprised of 262 hours of
Switchboard 1 audio with segmentations and transcripts provided by Mississippi State
University plus 1698 hours from the Fisher data collection with transcripts provided by LDC plus 15 hours of CallHome audio. We trained neural transducers with either conformer (10 or 12 layers) or DSSformer encoders (10 layers), feed-forward dimension of 512 and 8$\times$64-dimensional attention heads. All DSS layers use bidirectional kernels with a state space dimension $N$=96. Training was carried out on 4 A100 GPUs with an effective batch size of 128 for 20 epochs with a one cycle LR policy with a maximum learning rate of 5e-4. The other settings are the same as in~\ref{swb300}. In Table~\ref{swb2000-tab} we show a comparison of baseline conformer and DSSformer transducers with various $\mathbf{\Lambda}$ initializations. As can be seen, DSSformer encoders outperform the conformer counterparts and the best $\mathbf{\Lambda}$ initialization is the same as in~\ref{swb300}. For contrast, we also compare our results with the single best performing model on this task from~\cite{tuske2021limit} and note that we achieve a comparable performance on two out of three test sets.

\begin{table}[H]
\begin{center}
\setlength\tabcolsep{6pt} % default value: 6pt
\begin{tabular}{|l|c|c|c|c|c|} \hline
\multirow{2}{*}{Encoder} & \multicolumn{3}{|c|}{Hub5'00} & \multirow{2}{*}{Hub5'01} & \multirow{2}{*}{RT'03}\\ \cline{2-4}
         & swb  & ch    & avg  &      &    \\ \hline
Conformer (10L) &  5.2 & 8.5 & 6.9 & 7.6 & 7.8 \\
Conformer (12L) &  5.4 & 8.5 & 6.9 & 7.6 & 8.2\\ \hline
$\mathbf{\Lambda}$ HiPPO &  5.2 & 8.4 & 6.8 & 7.4 & 7.5\\ \hline
$\mathbf{\Lambda}$ S4D-Lin & 5.3 & 8.4 & 6.8 & 7.6 & 7.5\\ \hline 
$\lambda_n=-1+i\cdot n$ &  5.1 & 8.5 & 6.8 & 7.4 & 7.4\\ 
~~~+length perturb. & 5.2 & 8.2 & 6.7 & 7.2 & 7.5\\ \hline\hline
Conformer AED~\cite{tuske2021limit}& 4.8 & 8.0 & 6.4 & 7.3 & 7.5\\ \hline
\end{tabular}
\end{center}
\caption{\label{swb2000-tab} Recognition results for conformer (10 and 12 layers) and DSSformer transducers (10 layers) with different $\mathbf{\Lambda}$ initializations on Switchboard 2000 hours (Hub5'00, Hub5'01, RT'03 test sets). All decodings are done without external LM.}
\end{table}

\subsection{Experiments on MALACH 176 hours}
\label{malach}
Lastly, we test the proposed models on the public MALACH corpus~\cite{bhuvana03} (released by LDC as LDC2019S11) which consists of Holocaust testimonies collected by the
Survivors of the Shoah Visual History Foundation. The corpus is 16kHz audio broken
down into 674 conversations totaling 176 hours for training (880 hours after augmentation) and 8
conversations of 3.1 hours for testing. The collection consists of
unconstrained, natural speech filled with disfluencies, heavy accents,
age-related coarticulations, un-cued speaker and language switching,
and emotional speech, all of which present significant challenges for
current ASR systems. Because of this, the error rates reported are
significantly higher than for the previous corpora. We trained conformer and DSSformer transducers with the same feature extraction, architecture, DSS layer initialization and training recipe as in~\ref{swb300} without length perturbation and with S4D-Lin $\mathbf{\Lambda}$ initialization. In Table~\ref{malach-tab} we report results with and without external LM fusion where the LM is a 10 layer 512-dimensional transformerXL trained on 7.2M characters. Our results show a 7\% relative improvement in WER over the previous best hybrid LSTM approach.

\begin{table}[H]
\begin{center}
\begin{tabular}{|l|l|l|c|c|} \hline
Work & Model & Encoder & LM & WER\\ \hline
\cite{bhuvana03} & HMM & GMM & n-gram & 32.1\\ \hline
\multirow{2}{*}{\cite{picheny2019challenging}} & \multirow{2}{*}{HMM} & \multirow{2}{*}{LSTM} & n-gram & 23.9\\ 
                                         &                        &                          & LSTM & 21.7 \\ \hline
\multirow{3}{*}{Ours}                    & \multirow{3}{*}{RNN-T} & Conformer & -- & 21.5\\ \cline{3-5}
                                         &                        & \multirow{2}{*}{DSSformer} & -- & 20.9\\
                                         &                        &                            & Trafo & 20.2\\ \hline
\end{tabular}
\end{center}
\caption{\label{malach-tab} Performance comparison of conformer and DSSformer transducer with other single-model approaches from the literature on MALACH 176 hours.}
\end{table}

\section{Discussion}
\label{con-sec}
Diagonal state space models are a promising alternative to temporal convolutions with fixed-length kernels for ASR when used in a conformer-style architecture. We attribute their success to the connection with function approximation theory and to the interpretability of their parameters. In future work we will investigate better ways of integrating DSS layers with self-attention and feedforward modules as opposed to simply using them as a drop-in replacement for the depthwise convolutions in conformer. For example, the DSS mixing matrix can be combined with the second pointwise convolution which will simplify the overall architecture. Another avenue of research is to improve the initialization for the real parts of the eigenvalues of the state transition matrices and possibly keep the $\mathbf{\Lambda}$s fixed during training which will reduce the number of free parameters. Lastly, we plan to study the effectiveness of DSS for other end-to-end ASR modeling approaches.

\bibliographystyle{IEEEbib}
\bibliography{icassp2023}
\end{document}